\documentclass{article}
\usepackage[utf8]{inputenc}
\usepackage{balance}  % to better equalize the last page
\usepackage{graphicx}
\usepackage[T1]{fontenc}
\usepackage{txfonts}
\usepackage{mathptmx}
\usepackage{textcomp}
\usepackage{microtype}
\usepackage{ccicons}
\usepackage{url}

\begin{document}

\title{Towards Understanding Connections between Security/Privacy Attitudes and Unlock Authentication\thanks{Article to appear at the Workshop on Usable Security (USEC), Feb. 2018}}

% \author{\IEEEauthorblockN{Adam J. Aviv}
% \IEEEauthorblockA{United States Naval Academy\\
% aviv@usna.edu}
% \and
% \IEEEauthorblockN{Ravi Kuber}
% \IEEEauthorblockA{University of Maryland, Baltimore County\\
% rkuber@umbc.edu}
% }

\author{Adam J. Aviv\\
  United States NAval Academy\\
  {\tt aviv@usna.edu}
  \and
  Ravi Kuber\\
  University of Maryland, Baltimore County\\
  rkuber@umbc.edu
}

% \IEEEoverridecommandlockouts
% \makeatletter\def\@IEEEpubidpullup{9\baselineskip}\makeatother
% \IEEEpubid{\parbox{\columnwidth}{Permission to freely reproduce all or part
%     of this paper for noncommercial purposes is granted provided that
%     copies bear this notice and the full citation on the first
%     page. Reproduction for commercial purposes is strictly prohibited
%     without the prior written consent of the Internet Society, the
%     first-named author (for reproduction of an entire paper only), and
%     the author's employer if the paper was prepared within the scope
%     of employment.  \\
%     USEC '18, 18 February 2018, San Diego, CA, USA\\
%     Copyright 2018 Internet Society, ISBN 1-891562-42-8\\
%     http://dx.doi.org/10.14722/usec.2018.23xxx
% }

%\hspace{\columnsep}\makebox[\columnwidth]{}}

\maketitle

\begin{abstract}

In this study, we examine the ways in which user attitudes towards privacy and security relating to mobile devices and the data stored thereon may impact the strength of unlock authentication, focusing on Android's graphical unlock patterns. We conducted an online study with Amazon Mechanical Turk ($N=750$) using self-reported unlock authentication choices, as well as Likert scale agreement/disagreement responses to a set of seven privacy/security prompts. We then analyzed the responses in multiple dimensions, including a straight average of the Likert responses as well as using Principle Component Analysis to expose latent factors. We found that responses to two of the seven questions proved relevant and significant. These two questions considered attitudes towards general concern for data stored on mobile devices, and attitudes towards concerns for unauthorized access by {\em known} actors.  Unfortunately, larger conclusions cannot be drawn on the efficacy of the broader set of questions for exposing connections between unlock authentication strength (Pearson Rank $r=-0.08$, $p<0.1$). However, both of our factor solutions exposed differences in responses for demographics groups, including age, gender, and residence type. The findings of this study suggests that there is likely a link between perceptions of privacy/security on mobile devices and the perceived threats therein, but more research is needed, particularly on developing better survey and measurement techniques of privacy/security attitudes that relate to mobile devices specifically.

\end{abstract}

%%% Local Variables:
%%% mode: latex
%%% TeX-master: "main"
%%% End:

% \category{K.6.5}{Security and Protection}{Authentication}
% \category{H.5.2}{User Interfaces}{Evaluation/methodology}
% \category{K.4.1}{Computers and Society}{Privacy}

\section{Introduction}

In order to protect sensitive data stored on or accessed via a mobile
device, such as a smartphone, knowledge-based authentication methods
(e.g. use of PINs, stroke-based patterns) or biometric methods
(e.g. fingerprints) can be used to verify the identity of the user,
thereby enabling the device to be unlocked.  Research suggests that
attitudes to securing mobile devices are known to differ among
users~\cite{harbach2014sa, harbach2016anatomy, egelman2014readytolock,
  harbach2016keep,malkin2017anatomy,albaryam2017better,mare2016study}. Although locking behavior is often practiced,
annoyances have been found with the
process~\cite{harbach2014sa,albaryam2017better,mare2016study}. While not averse to locking a mobile
device, Egelman et al.~\cite{egelman2014readytolock} identified lack
of motivation as one factor deterring participants in their study from
securing their technologies.  Calculated decisions were made by some
participants, depending on the perceived effort required to lock
devices.  Deterrents have also been attributed in part to the absence
of threats to security, along with negative perceptions of the
security of lock screens~\cite{harbach2016anatomy}.

For purposes of unlocking a mobile device, many Android users have the option of using
stroke-based patterns which can be entered via the interface.  The user is required to connect a sub-set of four or more pre-selected contact points arranged in a 3x3 grid in the correct
sequence to authenticate entry.  However, studies have shown that in practice,
only a small number of distinct patterns are selected from the 389,112 choices
available~\cite{aviv2010smudge}.  The majority of these user-selected patterns
can be easily guessed with roughly the same difficulty as guessing random
3-digit PINs~\cite{aviv2015bigger,uellenbeck}.  While the threat of observer and
smudge attacks exist when using a locking mechanism, the benefits often outweigh
the impact of losing sensitive data for potential abuse by third parties.  Prior work in the area has often focused on the factors impacting locking compliance or lack thereof~\cite{harbach2014sa, harbach2016anatomy, egelman2014readytolock,
  harbach2016keep}. Although limited in number, studies have been conducted examining security and privacy attitudes associated with locking
behavior~\cite{albaryam2017better,mare2016study}. However, research has yet to be undertaken to directly connect security/privacy attitudes with passcode strength. %  While studies also assume that most users unlock their phones to
% perform mobile tasks (e.g. to check messages etc), comparatively less is also
% known about how users interact with their devices when not locking
% them~\cite{Hintze2014}.

In this paper, we describe a study examining the ways in which users select
authentication stimuli, notably Android's graphical password system or ``pattern
unlock'' based on their attitudes towards a set of security and privacy prompts
concerning information privacy on mobile devices. We collected and analyzed a
large dataset ($n=750$) of self-reported Android unlock patterns/statistics
regarding pattern choice and correlated that data with agreement/disagreement
responses to a set of privacy prompts related to the privacy of data stored on
mobile devices.

The goals of this study are twofold: first we hypothesized that there is a
relationship between privacy attitudes regarding mobile devices and the security
of the authentication stimuli selected (H1), and second we hypothesized that
there may be demographic factors (e.g., age, location of residence) that may
affect security and privacy responses (H2). As our questionnaire is short in
design, the hope is that these prompts can be used as starting point to design
more advanced measurement techniques, or more directly, a subset of these
questions could be used at device set-up time (or during a security audit) to
better understand user motivation and potentially nudge users towards secure
choices.

We conducted per-prompt analysis, an analysis of averages of the scores, and
used principal component analysis (PCA) to expose any latent factors in the
data.  The PCA analysis explained 55\% of the variance for a two factor solution
with high internal consistency for each factor solution (Cronbach's alpha
$>0.7$) and low correlation (likely independence) between the factors (Spearman
Rank $\rho=0.09$, $p<0.05$). The two different latent factors identified
different aspects of participant response, particularly around aspects of {\em
  present vs. external} threats to security/privacy. A similar dichotomy was found
when identifying connections to unlock authentication strength.  

The overall results, using the entirety of the questions to predict users with
weaker unlock strengths, are unfortunately {\em inconclusive} as the correlation
effect is small (Pearson Rank $r=-0.08$, $p<0.1$) with a weak effective
size. We argue, though, that the results are encouraging for the potential of
such a link between security/privacy attitude and unlock authentication may be
present, if more advanced measurement methods were developed.

The results of the survey do provide a path forward in identifying better honed
measurement methods when analyzing the response to individual privacy/security
prompts. Two such prompts had statistically significant differences with regard
to the reported strength of the authentication patterns. In particular, an
agreement with general concern for privacy of data stored on mobile devices and
concern of unauthorized access from known actors (e.g., friends and family) led
to stronger authentication choices $(p<0.05)$.

In analyzing other relationships between the participants and the
privacy prompts, we show that our two factor solutions found using PCA
exposes interesting demographic differences in our responses. In
particular, Factor 1 solution correlated more with the strength of the
reported results, our second factor solution, Factor 2, better
separated the respondent's attitude with respect to their age and
willingness to provide a pattern or statistics about those patterns
for the experiment.

To summarize, the paper makes the following contributions and findings:
\begin{itemize}
\item An analysis of the efficacy of a short privacy/security questionnaire
  regarding threats to mobile devices.
\item Exposing latent factors via PCA that describe clear difference in
  participants between present and external threats to mobile device security.
\item Demonstrating correlations between latent factors of the survey with
  demographic information, particularly age and location.
\item Identifying the connection of individual questions to unlock
  authentication strength, but inconclusive results for the entirety
  of the prompts.
\end{itemize}

Based on these results, we conclude that while we suspect that a
relationship between authentication choice and security/privacy attitudes of
mobile devices seems probable, we were not able to find enough evidence here to
draw strong conclusions. Further study is needed, and we show a number of
promising directions regarding measurement methods that could be applied to
expose this link. In particular, there are encouraging findings with respect to
specific prompts, as well as when applying factor analysis to the entire set of
privacy prompts. This suggests that using a short survey to better understand
user perceptions regarding privacy on mobile devices could be possible and would
be a useful tool in directing users in a personalized way, based upon their own
security and privacy perceptions, towards more secure choices.

% We find that
% attitudes toward privacy on mobile devices relatively lean towards
% more agreement with the privacy prompts. However, those with more
% privacy sensitivity tend to select stronger authentication, and
% stronger attitudes towards general information privacy as well as fear
% of known actors (e.g., friends and family) have the largest impact on
% users selecting stronger authentication.

%%% Local Variables:
%%% mode: latex
%%% TeX-master: "main"
%%% End:

\section{Related Work}
Researchers have aimed to optimize both security and usability
for mobile unlocking mechanisms, through extensive studies examining
both attitudes and usage behaviors relating to securing mobile technologies.  In terms of
attitudes, mobile device users have been found to express concerns
about security~\cite{kurkovsky2010digital} and make clear demands for
protection against unauthorized use~\cite{clarke2005authentication}.
However, difficulties have been experienced when complying with general security guidance, such as
updating passcodes~\cite{clarke2005authentication}, even if users are ‘tech
savvy’~\cite{kurkovsky2010digital}. Research suggests that a number of
mobile device users prefer not to lock their
devices~\cite{egelman2014readytolock}.  Reasons for this include
concerns about emergency personnel not being able to identify them,
not having their devices returned if lost, and not believing they had
any data worth protecting.  Lack of motivation and awareness were also
found to be reasons behind lack of compliance in a study by Qiu et
al.~\cite{qiu2016advancing}  The usability challenges (i.e., the time taken to use locking screens) were identified by Malkin et al.~\cite{malkin2017anatomy} as deterrents for locking. When examining perceptions in more
detail, Egelman et al.~\cite{egelman2014readytolock} identified a
strong correlation between use of security features and risk
perceptions, which indicates rational behavior.  The researchers also
observed that most users likely underestimate the extent to which data
stored on their smartphones pervades their identities, online and
offline.  Hayashi et al.~\cite{hayashi2012goldilocks} found that most
of the participants in their study wanted at least half of their
applications to be accessible without requiring an unlock code, which
according to Egelman et al.~\cite{egelman2014readytolock}, suggests opportunities for improving current locking mechanisms.

Studies have examined a range of factors which may play a role in
impacting privacy and security attitudes.  For example, Benensen et
al.~\cite{Benenson2013} conducted a study to identify whether mobile
device users utilizing different operating systems had varying
perceptions of security. Findings from their study showed that 20\% of
Android users and 5\% of iOS users were privacy aware.  Mare et al. \cite{mare2016study} examined the impact of age, among other factors, on attitudes to privacy and security.  The researchers identified that teenage users and not just adults were interested in these areas.  However, teenage users were thought to have less useful understandings of how to achieve it. 

Studies have been conducted to investigate locking behaviors by users from eight countries (e.g. \cite{harbach2016keep,malkin2017anatomy}).
Findings revealed that level of protection varies by location.  Participants in most non-U.S. countries were
between 31\% and 76\% more likely than Americans to have a secure lock
screen.  Age and sex were also thought to play a role in determining
locking behavior.  Reasons for lack of compliance appeared to be
similar across participants from most countries examined \cite{harbach2016keep}.  Egelman and
Peer~\cite{egelman2015predicting} examined the impact of individual
differences on privacy and security attitudes.  The five factor model,
often used to examine the impact of these differences on decision
making, was found to be a weak predictor of privacy preferences and
behaviors.  The researchers identified that decision-making style and
risk-taking attitudes are strong predictors of privacy attitude.

Qiu et al.~\cite{qiu2016advancing} have focused on recording and
analyzing locking behaviors among mobile device users.  These include
the length of unlocking and interaction sessions with the devices,
outcomes (failed/succeeded) of the unlocking attempts, type of locking
mechanism (e.g. manually or through auto-lock), participants'
attitudes of sharing their devices with other people, device locations
and information gathered from accelerometers/gyroscopes, the model of the device and its screen size.
Although the study has yet to be completed, the findings would offer
interesting insights to interface designers aiming to develop security
solutions to better meet the needs of mobile device users.  Hintze et
al.~\cite{Hintze2014} also examined usage behavior when using mobile
devices.  Findings showed that a noteworthy amount of users do not
lock their devices after usage, increasing the possibility of
malicious interaction with the device when considering unlocked usage,
as devices stay unlocked for a short period afterwards.  

In terms of challenging perceptions of risk, concerns and attitudes to screen locking, Albayram et al. \cite{albaryam2017better} presented a video to participants based on 'fear appeal' highlighting the potential consequences associated with poor locking compliance. The video was found to significantly affect levels of perceived severity, vulnerability, and response efficacy.  The researchers suggest that risk communication can effectively change risk perception, which can be a key to promoting secure behavior such as the use of a secure screen lock mechanism on smartphones.

%%% Local Variables:
%%% mode: latex
%%% TeX-master: "main"
%%% End:

\begin{table*}[t]
  \centering
%  \small
  \resizebox{\linewidth}{!}{
  \begin{tabular}{r | c  c | c c  c | c c }
     & {\em Reported} & {\em Reported} & & & &\\
    $n=750$        & {\em Statistics} & {\em Pattern} & {\em Urban} & {\em Suburban} & {\em Rural} & {\em Male} & {\em Female} \\
    \hline
    Participants &  307 & 443 & 278 & 330 & 142 & 407 & 343 \\
    \hline
    18-24 & 58 & 148 & 83 & 90 & 33 & 123 & 83\\
    25-34 & 134 & 228 & 143 & 154 & 65 & 199 & 163\\
    35-44 & 85 & 52 & 41 & 66 & 30 & 67 & 70\\
    45-54 & 24 & 13 & 8 & 18 & 11 & 17 & 20\\
    55-64 & 5 & 2 & 3 & 2 & 2 & 1 & 6\\
    65+ & 1 & 0 & 0 & 0 & 1 & 0 & 1\\
    \hline
    Urban & 119 & 159 &  &  &  & 170 & 108\\
    Suburban & 133 & 197 &  &  &  & 173 & 157\\
    Rural & 55 & 87 &  &  &  & 64 & 78\\
    \hline
    Male & 153 & 254 & 170 & 173 & 64 &  & \\
    Female & 154 & 189 & 108 & 157 & 78 &  & \\
  \end{tabular}}
  \caption{Participant demographics for those that only reported
    statistics on their authentication pattern and those that reported their
    authentication pattern.}
  \label{tab:dem}
\end{table*}

\section{Methodology}
The goal of this research was to determine if security and privacy attitudes
towards data stored on mobile devices impact the strength of chosen
passcodes. Our primary result involved the analysis of participants'
agreement/disagreement with statements regarding information security/privacy of
mobile phones (on a Likert scale).  This type of approach has been used by other
researchers in prior
work~\cite{caine2009understanding,kwasny2008privacy,malheiros2013fairly,
  benenson2014user, chanchary2015user}). Some of those participants also
self-reported their mobile device unlock passcode (an Android graphical pattern)
in addition to reporting their agreement/disagreement with the privacy
statements, allowing us to compare their privacy sentiment with regards to their
mobile device and the relative strength of their passcodes.

To measure the strength of the reported passcodes, we use password guessability
metrics~\cite{bonneau2012science,mazurek2013measuring,kelley2012guess} and
particularly the guessing methods developed in related
work~\cite{aviv2015bigger, uellenbeck} trained using non-overlapping data of
user-generated passcodes obtained from the same related
work~\cite{uellenbeck,loge2016user,vz2016onquant}. All the data collection
methods were reviewed and approved by our institutions' IRBs.

\subsection{Self Reporting Passcodes}
This study builds upon the analysis and data collection conducted previously by
Aviv et al.~\cite{aviv2015bigger,aviv2014understanding}. The study was
conducted on Amazon Mechanical Turk and participants who self-identified as
Android mobile device users were recruited with the following prompt:
\begin{quote}
  You will be asked to (optionally) self-report what your Android
  unlock pattern is, or report statistics about your pattern. If you
  do not use the Android unlock pattern, please do not complete the
  HIT.
\end{quote}
Participants entering the survey were aware of the nature of the study
and what was going to be asked of them. This may have provided a bias to the
results as those {\em opting-in} may have been less privacy conscious then
those {\em opting-out}; however, the option to {\em not} report a
pattern and still complete the survey allowed for more privacy
conscious users to participate. The statistics that participants
reported about their patterns included the start contact point and whether any
of a pre-selected set of tri-grams were present, i.e., a connection of
three contact points, but were not required to reveal their precise
pattern. Both those who reported their actual pattern or statistics
were compensated equally, \$0.75, for their participation and fully
completing the survey.

Once the HIT was accepted, the structure of the survey was as follows:
\begin{enumerate}
\item Informed Consent: Participants were made aware of the intent of the
  survey, the data collection and protection procedures, and the kinds of
  information that would be collected. 
\item Demographic Information: Participants answered background questions about
  their age, gender, and place of residence, either rural, suburban, or
  urban. Participants were limited to U.S. areas using Amazon Mechanical Turk's
  settings.
\item Pattern/Stats: Participants were given the option to either report
  statistics about their Android unlock pattern or report their actual pattern.
\item Privacy Prompts: Participants were asked to rate their
  agreement/disagreement with the privacy prompts (see below for specific
  prompts).
\item Confirm/Attention Tests: At the end of the survey, participants were asked to
  re-enter their pattern (or stats) and answer a question about providing
  truthful responses. Participant data was removed if the participant failed to confirm
  the same pattern/stats or answered negatively to the truthfulness
  question. Seven participants were excluded for this reason.
\end{enumerate}

Further, participants were required to complete the entirety of the survey on an
Android device. The survey, including pattern entry screens were written using a
combination of JavaScript, PHP, and HTML, and run completely within a browser on
a mobile device. Once the HIT was accepted, the participants were provided with a link
to visit on their Android device's browser window. To ensure that participants
used an Android device, the {\tt user-agent} header was checked. While the {\tt
  user-agent} header can be spoofed, it is likely more burden than it is worth
given the compensation for the survey of \$0.75. Participants who did not own an
Android device should have returned the HIT in order to complete other tasks.

The demographics of the study are presented in Table~\ref{tab:dem}, and in total
750 individuals participated in the study, 443 of which also self-reported their
authentication passcode for their Android unlock pattern.

\subsection{Security/Privacy Prompts}
Each participant, regardless of if they self-reported a unlock
pattern, were provided with the following prompts/questions which they responded disagreement/agreement with on the following
scale: ``Strongly Disagree,'' ``Disagree,'' ``Neither Agree nor
Disagree,'' ``Agree,'' or ``Strongly Agree.''
\begin{itemize}

\item {\bf Q1}: I am concerned about the privacy of the data stored on
  my mobile device(s).

\item {\bf Q2}: I am more concerned about the privacy of the
  information stored on my mobile device as compared to my laptop or
  desktop.

\item {\bf Q3}: I am more concerned about the privacy of the
  information stored on my mobile device as compared to information
  stored online.

\item {\bf Q4}: I am concerned about identity theft.

\item {\bf Q5}: I am concerned about my privacy online.

\item {\bf Q6}: I am concerned about unauthorized access to my device
  by known actors (people I know, for example, friends, family, and my
  peers).

\item {\bf Q7}: I am concerned about unauthorized access to my device
  by unknown actors (people I don't know, for example, police,
  government, and corporations).
\end{itemize}
Based on the responses, a security/privacy attitude average can be calculated
by assigning a value of 1 to ``Strongly Disagree,'' 2 to
``Disagree,''3 to ``Neither Agree nor Disagree'', 4 to ``Agree'', and
5 to ``Strongly Agree'' that related to the general privacy attitude
of a participant as it relates to information stored on mobile
devices.  For those participants that also self-reported their
authentication passcode, we next analyzed the strength of that passcode
as it related to their security/privacy attitudes. 

While other published scales have been used by researchers to determine
perceptions/attitudes towards security/privacy, such as the ones described by
Buchanan et al.~\cite{buchanan2007}, Malhotra et
al.~\cite{malhotra2004internet}, Kumaraguru and Cranor
\cite{kumaraguru2005privacy}-(Westin Primary Segmentation index), and others,
some parts of these scales are out of date, not focusing on mobile devices (and aspects of
unlock authentication) or were deemed to be prohibitively long for a participant
to take on a mobile device. Instead, we developed our prompts from prior
published privacy scales for computing,
generally~\cite{caine2009understanding,kwasny2008privacy}). We do not imply
that these prompts provide complete clarity of security/privacy
attitudes. However, they do expose some key aspects of the mental models of
users. For example, the difference between security/privacy of data stored on
the device vs. security/privacy of data stored online, as well as the difference
between unauthorized access by unknown actors and known actors. Further, as we
will discuss in the next section, we found acceptable internal consistency with
two different factors, similar to that of other published security/privacy
questionnaires.

%% num= 750
%% pat= 443

%% ages
%% 18 206 148
%% 25 362 228
%% 35 137 52
%% 45 37 13
%% 55 7 2
%% 65 1 0

%% loc
%% Urb 278 159
%% Sub 330 197
%% Rul 142 87

%% gender
%% Male 407 254
%% Female 343 189

%%% Local Variables:
%%% TeX-master: "main"
%%% End:

\begin{figure}[t!]
  \centering
  \includegraphics[width=1.0\linewidth]{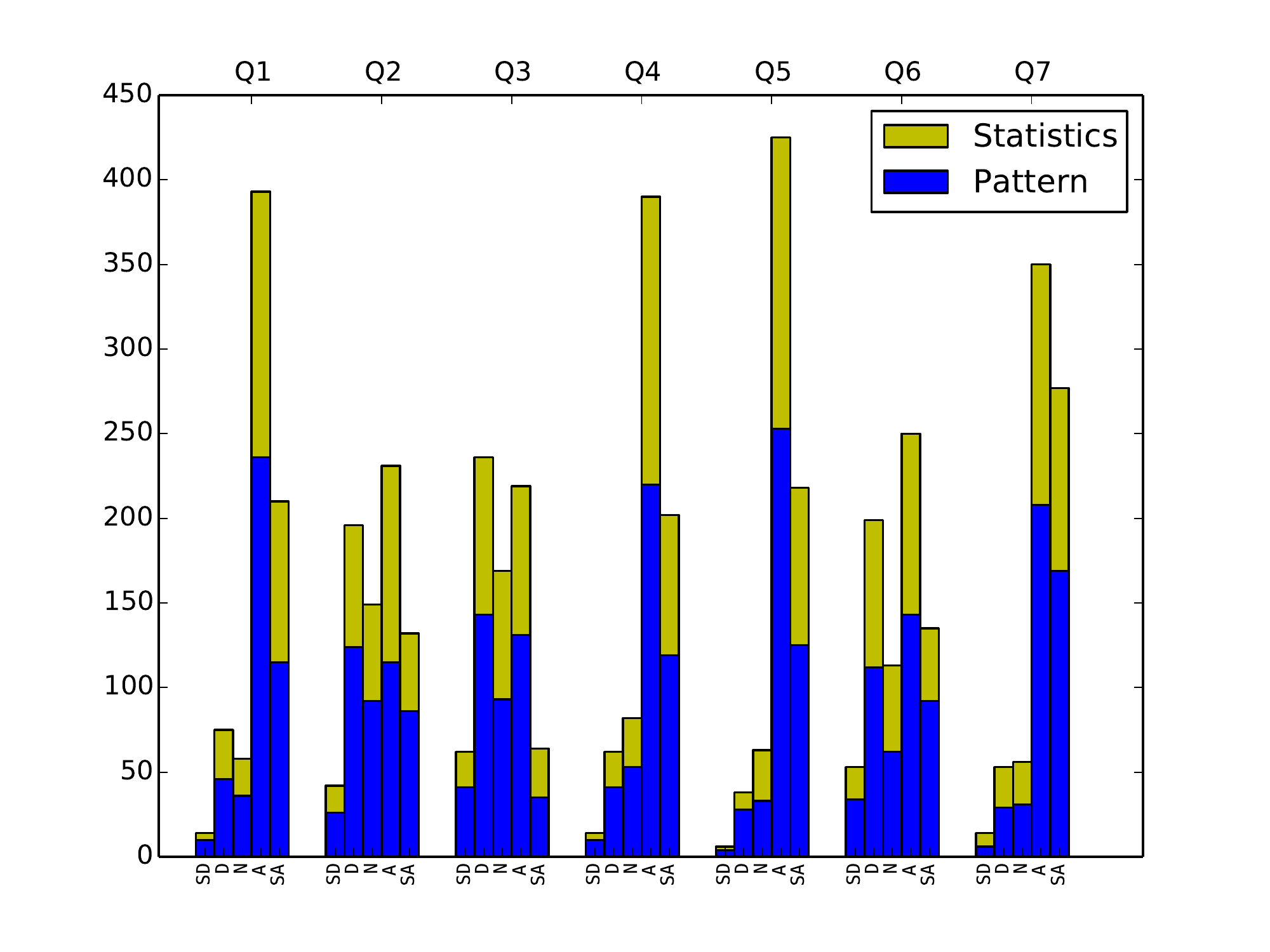}
  \caption{Magnitude of responses to each question prompt for those
    that reported statistics and those that reported patterns, where:
    SD is ``Strongly Disagree,'' D is ``Disagree,'' N is ``Neither
    Disagree nor Agree,'' A is ``Agree,'' and SA is ``Strongly
    Agree.''}
  \label{fig:magnitude}
\end{figure}

\section{Identifying Latent Factors}
As a first step in the analysis, consider the results in
Figure~\ref{fig:magnitude} for a general overview of the responses to each of
the privacy prompts. Clearly observed is the fact that there are varied
distributions in the responses, by which some of the participants were strongly
in agreement for most of the questions, while other questions provided more
varied responses. The variance of the responses is not ideal as some questions
have mostly agreement or strong-agreement (Q1, Q4, Q5, and Q7), while others
show more spread variance (Q2, Q3, and Q6). To better identify the impact of all
the prompts, i.e., to find an appropriate weighting, we can apply a naive method
or a more advanced factor-based analysis.

The naive method for computing a quantitative privacy response into a continuous variable for each participant is
to simply take the average of each of the responses to each of the questions, on
a scale from 1 to 5, which we refer to as the {\em privacy average}. A privacy
average across the prompts of 5 would indicate strong agreement with all the
prompts, and a 1 would indicate strong disagreement. Values in-between would
indicate a more nuanced perspective, but likely shifted towards more
agreement or less agreement with the prompts. We apply this quantization in much
of analysis and justify it through principal component analysis to follow.

As a more advanced technique for determining appropriate weighting of the
questions in computing a continuous value, we use principal component analysis (PCA) to identify latent factors
in the responses and the associated question weights (loading) that lead to high
internal consistency in responses between questions. This is a common technique
for verifying privacy scales, and we follow the verification technique in prior
published work, particularly that of Buchanan et al.~\cite{buchanan2007}.

\begin{figure}
  \centering
  \includegraphics[width=0.9\linewidth]{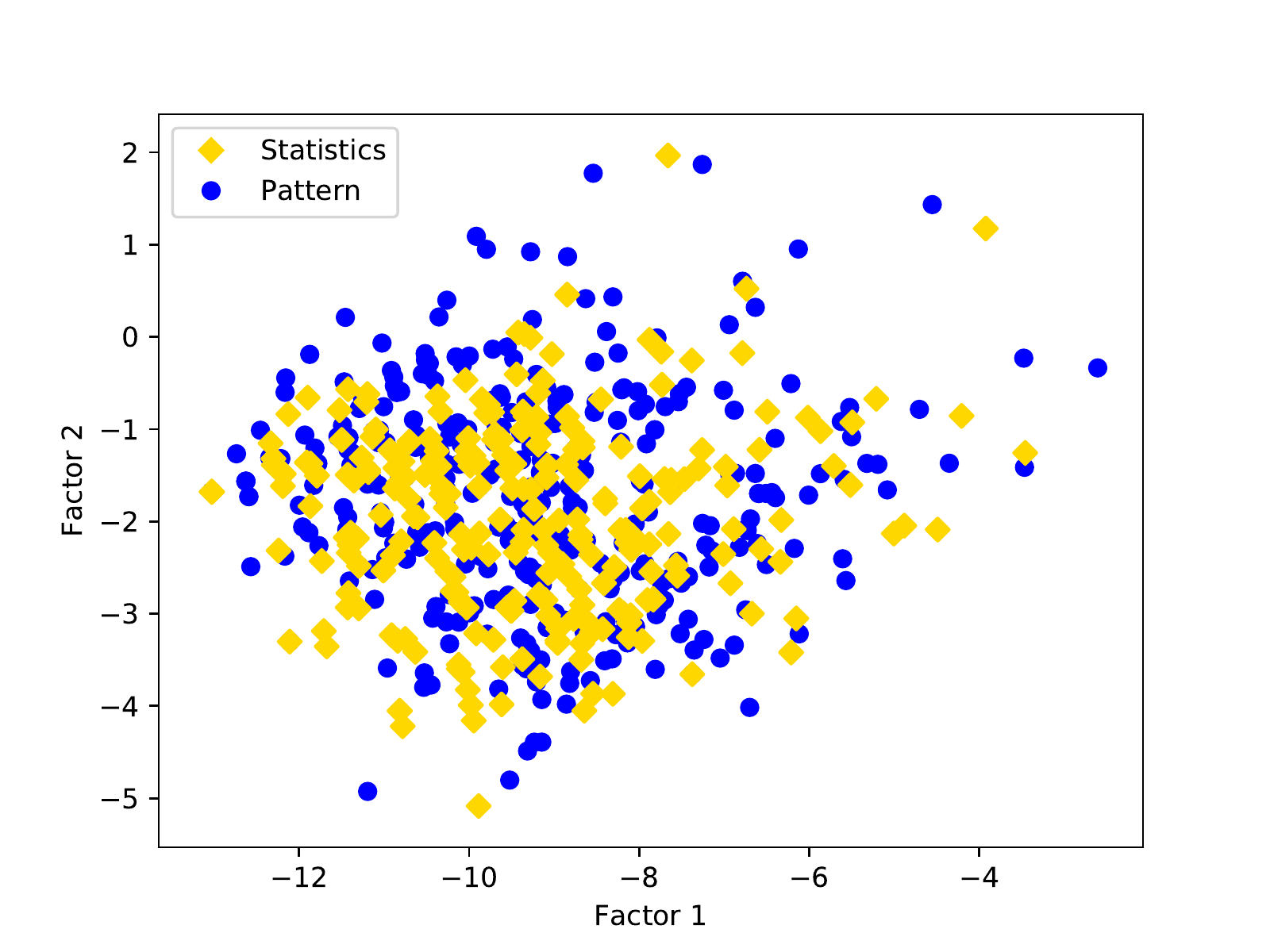}
  \caption{A scatter plot of two factor solution correlation. The two factors
    are weakly negatively correlated based on Spearman's rank correlation coefficient
    ($\rho=-0.088$, $p<0.05$).}
  \label{fig:factors}
\end{figure}

Just as in Buchanan et al.~\cite{buchanan2007}, we required a loading greater than 0.3 for
a Varimax analysis, which left us with a three-factor solution. Further we used
Saucier's criteria and then Cronbach's alpha $>0.7$ to determine factor purity,
after which we were left with two viable factors. The loadings of each of the
factors are presented in Table~\ref{tab:factors}. Factor 1 (F1) had a Cronbach
alpha of 0.74 and Factor 2 (F2) had a Cronbach alpha of 0.71. Comparing the two-factor solutions, we find that they are weakly correlated. Figure~\ref{fig:factors}
% * <eskchan@gmail.com> 2017-08-23T19:46:55.583Z:
%
% ^.
plots the factors in 2D space, and shows the weak negative correlation (Spearman
rank of $\rho=-0.09$, $p<0.05$), suggesting that each factor is exposing
different latent factors of the. As $\rho$ is quite small, these two factors are
exposing different aspects of the responses, and overall, 55\% of the variance
is explained by the two factor solution.

\begin{table}
\small
\centering
\begin{tabular}{l | c  c}
     & F1 & F2 \\
\hline
  Q1 & -0.411 & -0.245  \\
  Q2 & -0.435 &  \\
  Q3 & -0.401 & -0.114 \\
  Q4 & -0.290 & -0.411 \\
  Q5 & -0.282 & -0.255 \\
  Q6 & -0.458 &  0.812 \\
  Q7 & -0.328 & -0.176 \\
\end{tabular}
\caption{Weighting for the two factors found via PCA where 1 is ``Strongly Disagree'' and 5 is ``Strongly Agree''. The Cronbach $\alpha$ > 0.7 (acceptable) for these values}
\label{tab:factors}
\end{table}

Observing the loadings, F1 is very similar to the privacy average metric
described previously because each of the questions are loaded at nearly equal
weight, but the loading is all negative. Since the contribution of each of the
questions is significant and none are significantly greater than any other, it
suggests that using an average of the results provides a reasonable and
straightforward quantitative measure of privacy attitudes based on the question
prompts.

The F2 solution provides a very different loading. Question 2, which compares
the privacy of information stored on the mobile device compared to a laptop or
desktop, does not contribute to the solution based on the minimization criteria
requiring high internal consistency. While Q2 contributes highly in F1, it does
not provide any loading in F2. Instead, Q6 and Q4 have the largest impact, one
positive and one negative, respectively. 

These two questions, Q6 and Q4, highlight one of the big dividers we found in
this survey: {\em present vs. distant threats}. Q4 refers to the risk of
identity threat, which mostly occurs from unknown actors, while Q6 refers to the
threat of unauthorized access by familiar, known actors, like friends and
family. When the external threat factor is combined with the other negatively
loaded factors, which all deal with external threats to data security and
privacy, it is clear that F2 is picking up on differences between participants
fears of known vs. unknown actors. As we will show, F2 provides significant
differences in results when considering the willingness of demographic factors,
particularly surrounding age.

% Additionally, the significance of Q6 is even greater when we consider that it is
% absolute heighest loading in both of the factor solutions. As we will show, it
% also proves to be a significant divider with regard pattern strength. Similarly,
% Q1 also shows significant differences for pattern strength, however, its loading
% varies between factor solutions.

% In total, this

% Further, most questions have a negative loading, while Q6 (fear of known
% actors) has a positive loading. This suggests an interesting role for Question 6
% in our privacy scale as a good divider for participants privacy
% attitudes. \reminder{add some more here about what this weighting might mean}

%%% Local Variables:
%%% mode: latex
%%% TeX-master: "main"
%%% End:

\section{Guessability Metric}
For the strength metric of Android unlock patterns, we use
guessability~\cite{bonneau2012science, mazurek2013measuring,
  kelley2012guess}. Guessability is a metric that describes how many guesses
would an informed attacker require to guess a user's password in an offline
setting.

The guessability metric requires the creation of a guessing algorithm which
outputs a series of guesses in order of most likely used passwords to least
likely used passwords. Prior work in this space has provided a framework for
designing these algorithms~\cite{aviv2015bigger,uellenbeck} which require
training data. One advantage present for studying Android unlock patterns is
that, unlike text-based passwords where there are essentially an unbound number
of passwords, for Android unlock patterns there are finite number of patterns,
exactly 389,112~\cite{aviv2010smudge}. Thus, it is possible to develop a complete
ranking of patterns using a guesser.

The best known guessing algorithms~\cite{aviv2015bigger,uellenbeck} leverage the
properties of commonly selected patterns, such as the {\bf Z}- and {\bf
  M}-shaped patterns (and rotations/flips thereof), to construct a Markov model
based on transitions within the pattern grid space. Recall that Android unlock
patterns require connecting a set of 3x3 grid points in a continuous gesture
without repetition or avoidance. The order in which those grid points connect is
well modeled using a bi-gram transition matrix of likelihoods, where the
likelihoods are measured based on training data. Additional probabilities
considering the start point, end point and the length of the pattern can also be
modeled using similar measures. Once the model is constructed, all 389,112
possible patterns (enumerated using brute force, as in Aviv et. al~\cite{aviv2010smudge}) are analyzed using the Markov model, producing a likelihood
score. Ranking the patterns based on this likelihood score provides a
guessability metric where lower values indicate patterns that are more likely to
be guessed and higher values indicate patterns that are less likely to be
guessed.

We followed the Markov model construction from prior work by Aviv et
al.~\cite{aviv2015bigger} (particularly the pen-and-paper data). In addition to
training the model using the Android unlock patterns used in prior collections,
we also trained the Markov model on patterns provided via Uellenbeck et
al.~\cite{uellenbeck}. As a limitation, these models change over time as more
training data is available; however, repeated observations of user choice in Android
unlock patterns find that the set of common patterns appears to be
consistent~\cite{andriotis2016study, siadati2015fortifying,
  von2016quantifying,loge2016user,vz2016onquant}.

%%% Local Variables:
%%% mode: latex
%%% TeX-master: "main"
%%% End:

\section{Results}

\begin{table}[t!]
  \centering
  \small
  %\resizebox{\linewidth}{!}{%
  \begin{tabular}{ c | c c | c }
       & Reported Statics & Reported Pattern & $t$-test \\
    \hline
    Q1 & $\mu=3.90,\sigma=0.97$ & $\mu=4.01,\sigma=0.94$ & $p=0.135$ \\
    Q2 & $\mu=3.25,\sigma=1.22$ & $\mu=3.34,\sigma=1.14$ & $p=0.319$ \\
    Q3 & $\mu=2.95,\sigma=1.14$ & $\mu=3.04,\sigma=1.11$ & $p=0.284$ \\
    Q4 & $\mu=3.90,\sigma=0.98$ & $\mu=4.00,\sigma=0.87$ & $p=0.135$ \\
    Q5 & $\mu=4.05,\sigma=0.83$ & $\mu=4.12,\sigma=0.76$ & $p=0.266$ \\
    Q6 & $\mu=3.33,\sigma=1.27$ & $\mu=3.22,\sigma=1.18$ & $p=0.229$ \\
    Q7 & $\mu=4.14,\sigma=0.90$ & $\mu=4.04,\sigma=0.99$ & $p=0.137$ \\
    \hline
    avg & $\mu=3.65,\sigma=0.65$ & $\mu=3.68,\sigma=0.64$ & $p=0.474$ \\
    F1 & $\mu=-9.36,\sigma=1.75$ & $\mu=-9.44,\sigma=1.70$ & $p=0.521$ \\
    F2 & $\mu=-1.95,\sigma=1.08$ & $\mu=-2.12,\sigma=1.00$ & *$p=0.033$ \\
  \end{tabular}%}
  \caption{Security/Privacy average per question/prompt comparing those that
    reported statistics about their authentication to those that
    self-reported their actual authentication pattern. Data was normal
    ($p<0.01$) according to Anderson-Darling test. Performed MANOVA tests, differences across responses were {\em not} significant ($Z*=14.07,\alpha=0.056$). }

 \label{tab:prompt-stats:report}
\end{table}
    
\begin{table}[t!]
  \centering
  \small
  %\resizebox{\linewidth}{!}{%
  \begin{tabular}{ c | c c | c }
    & Male & Female & $t$-test \\
    \hline
    Q1 & $\mu=3.97,\sigma=1.00$ & $\mu=3.92,\sigma=0.91$ & $p=0.460$ \\
    Q2 & $\mu=3.23,\sigma=1.20$ & $\mu=3.35,\sigma=1.17$ & $p=0.163$ \\
    Q3 & $\mu=2.98,\sigma=1.16$ & $\mu=2.99,\sigma=1.10$ & $p=0.951$ \\
    Q4 & $\mu=3.90,\sigma=0.97$ & $\mu=3.98,\sigma=0.88$ & $p=0.239$ \\
    Q5 & $\mu=4.08,\sigma=0.82$ & $\mu=4.08,\sigma=0.78$ & $p=0.993$ \\
    Q6 & $\mu=3.31,\sigma=1.22$ & $\mu=3.26,\sigma=1.25$ & $p=0.540$ \\
    Q7 & $\mu=4.15,\sigma=0.90$ & $\mu=4.03,\sigma=0.98$ & $p=0.069$ \\
    \hline
    avg & $\mu=3.66,\sigma=0.63$ & $\mu=3.66,\sigma=0.66$ & $p=0.940$ \\
    F1  & $\mu=-9.39,\sigma=1.69$ & $\mu=-9.39,\sigma=1.77$ & $p=0.941$ \\
    F2  & $\mu=-2.00,\sigma=1.07$ & $\mu=-2.04,\sigma=1.04$ & $p=0.569$ \\
  \end{tabular}%}
  \caption{Security/Privacy average per question/prompt comparing male and
    female responses. Data was normal ($p<0.01$) according to
    Anderson-Darling test. Performed MANOVA tests, differences across responses were {\em not} significant ($Z*=14.07,\alpha=0.093$).}
    \label{tab:prompt-stats:gender}
\end{table}

\begin{table*}[t!]
\small
  \centering
%  \resizebox{\linewidth}{!}{%
    \begin{tabular}{ c | c c c | c }
      &  Urban & Sub-Urban & Rural & ANOVA \\
      \hline
      Q1 & $\mu=4.04,\sigma=1.00$ & $\mu=3.92,\sigma=0.91$ & $\mu=3.85,\sigma=0.99$ & $p=0.114$\\
      Q2 & $\mu=3.37,\sigma=1.18$ & $\mu=3.15,\sigma=1.18$ & $\mu=3.42,\sigma=1.20$ & *$p=0.024$\\
      Q3 & $\mu=3.07,\sigma=1.20$ & $\mu=2.86,\sigma=1.05$ & $\mu=3.09,\sigma=1.15$ & $p=0.032$\\
      Q4 & $\mu=3.94,\sigma=1.01$ & $\mu=3.94,\sigma=0.85$ & $\mu=3.94,\sigma=0.96$ & $p=1.000$\\
      Q5 & $\mu=4.13,\sigma=0.84$ & $\mu=4.04,\sigma=0.74$ & $\mu=4.09,\sigma=0.86$ & $p=0.411$\\
      Q6 & $\mu=3.43,\sigma=1.23$ & $\mu=3.15,\sigma=1.22$ & $\mu=3.32,\sigma=1.23$ & *$p=0.017$\\
      Q7 & $\mu=4.13,\sigma=0.97$ & $\mu=4.12,\sigma=0.86$ & $\mu=3.98,\sigma=1.05$ & $p=0.249$\\
      \hline
      avg & $\mu=3.73,\sigma=0.70$ & $\mu=3.60,\sigma=0.56$ & $\mu=3.67,\sigma=0.71$ & *$p=0.04$\\
      F1 & $\mu=-9.59,\sigma=1.86$ & $\mu=-9.20,\sigma=1.51$ & $\mu=-9.43,\sigma=1.89$ & *$p=0.02$\\
      F2 & $\mu=-1.95,\sigma=1.07$ & $\mu=-2.10,\sigma=1.04$ & $\mu=-1.96,\sigma=1.05$ & $p=0.09$\\
      
    \end{tabular}
    %}
    \caption{Security/Privacy average per question/prompt comparing locales. Data
      was normal ($p<0.01$) according to Anderson-Darling test. Using MANOVA indicates that there are significant differences across responses ($Z*=23.68, p=0.012$), and post-hoc analysis (one vs. all) suggests that responses from Urban participants were significantly different. Using a Bonferoni correction, Q2 and Q6 show significant different responses within questions. F1 comparisons also suggest significant differences.}
    \label{tab:prompt-stats:place}
\end{table*}

\begin{table*}[t!]
\small
  \centering
  \resizebox{\linewidth}{!}{%
    \begin{tabular}{ c | c c c c | c }
      &  18-24 & 25-34 & 35-44 & 45-54 & ANOVA\\
      \hline
      Q1 & $\mu=3.89,\sigma=1.02$ & $\mu=3.91,\sigma=0.99$ & $\mu=4.05,\sigma=0.83$ & $\mu=4.19,\sigma=0.65$ & $p=0.407$\\
      Q2 & $\mu=3.21,\sigma=1.25$ & $\mu=3.30,\sigma=1.18$ & $\mu=3.45,\sigma=1.13$ & $\mu=3.08,\sigma=1.07$ & $p=0.302$\\
      Q3 & $\mu=2.88,\sigma=1.19$ & $\mu=2.92,\sigma=1.13$ & $\mu=3.26,\sigma=1.01$ & $\mu=3.24,\sigma=1.02$ & *$p=0.017$\\
      Q4 & $\mu=3.87,\sigma=1.03$ & $\mu=3.96,\sigma=0.94$ & $\mu=3.97,\sigma=0.78$ & $\mu=4.03,\sigma=0.75$ & $p=0.814$\\
      Q5 & $\mu=4.01,\sigma=0.90$ & $\mu=4.09,\sigma=0.80$ & $\mu=4.15,\sigma=0.65$ & $\mu=4.14,\sigma=0.58$ & $p=0.439$\\
      Q6 & $\mu=3.51,\sigma=1.24$ & $\mu=3.25,\sigma=1.25$ & $\mu=3.23,\sigma=1.15$ & $\mu=2.84,\sigma=1.05$ & *$p=0.003$\\
      Q7 & $\mu=4.22,\sigma=0.87$ & $\mu=4.02,\sigma=1.00$ & $\mu=4.14,\sigma=0.86$ & $\mu=4.11,\sigma=0.86$ & $p=0.080$\\
      \hline
      avg & $\mu=3.66,\sigma=0.69$ & $\mu=3.64,\sigma=0.65$ & $\mu=3.75,\sigma=0.56$ & $\mu=3.66,\sigma=0.50$ & $p=0.385$ \\
      F1 & $\mu=-9.40,\sigma=1.85$ & $\mu=-9.33,\sigma=1.74$ & $\mu=-9.63,\sigma=1.50$ & $\mu=-9.34,\sigma=1.37$ & $p=0.477$\\
      F2 & $\mu=-1.79,\sigma=0.96$ & $\mu=-2.03,\sigma=1.13$ & $\mu=-2.16,\sigma=0.97$ & $\mu=-2.52,\sigma=0.85$ & *$p<0.001$\\
    \end{tabular}
    }
    \caption{ Security/Privacy average per question/prompt comparing Age. Age ranges for 55+ were excluded to limited sample size. Data was normal ($p<0.01$) according to Anderson-Darling test. Using MANOVA indicates that there are significant differences across responses ($Z*=23.68,p=0.012$), post-hoc analysis (one vs. all) suggest that responses from 18-24 and 45-54 were significantly different. Using a Bonferoni correction, Q3 and Q6 show significant difference in responses. Additionally, F2 captures differences in age well, again, 18-24 and 45-54 showing the most difference. }

    \label{tab:prompt-stats:age}
\end{table*}

\subsection{Demographic Differences in Responses}
Figure~\ref{fig:magnitude} provides the general overview of the
magnitude of responses for participants to the security/privacy prompts. The
figure separates by two groups: those willing to self-report their
authentication passcode (or pattern) and those unwilling. The
willingness to reveal this information could also be considered a
sensitivity regarding security and privacy, and we present the statistical analysis
in Table~\ref{tab:prompt-stats:report}.

An omnibus MANOVA test showed no signficant differences between these
groups. Additionally, the Factor 1 (F1) solution also showed no statistical
differences between the responses of participants who were or were not willing
to provide a pattern for analysis. Interestingly, the Factor 2 (F2) solution
does show a significant difference that is not present in prior tests. This
result suggests that F2 is capturing some more difference between the
willingness of those to share information about their authenticator than what is
simply captured under an average of the responses.

While we found no significant differences with respect to gender (see
Table~\ref{tab:prompt-stats:gender}), we did observe differences within other
demographic groupings. The reported location of participants (Urban, Sub-Urban,
vs. Rural) showed significant differences for the factor results and under
MANOVA tests ($p=0.012$); post-hoc, one-vs-all analysis suggests that Urabn
respondents differed significantly compared to other groupings (see
Table~\ref{tab:prompt-stats:place}). This is also captured using the F1
weighting and using the average of responses, as expected given the generally
smooth weighting of F1. In analyzing the questions, using a Bonferoni
correction, Q2 (more concern about security/privacy on mobile
vs. desktop/laptop) and Q6 (more concern about known actors vs. unknown actors)
showed the most differences for these groups. While we exclude Q3 (more concern
about privacy on mobile vs. online) due to the correction, it is worth noting
that it does seem to have some impact on differentiating respondents and is
highly weighted under F1, so is considered well in that comparison.

When comparing age groups (see Table~\ref{tab:prompt-stats:age}), again we see
that particular questions expose significant differences between the groups,
supported under MANOVA analysis ($p=0.012$). Post-hoc MANOVA analysis, using a
one-vs.all method, suggest that 18-24 and 45-54 age groups were significant
different in responses to other respondents. In particular, for older
participants (45-54) there is more agreement with Q3 regarding more concern for the
security/privacy of data stored on mobile devices as compared to data stored
online, but younger participants (18-24), however, show more disagreement, suggesting
that they may be more concerned with their online footprint. Further, there is a
divide regarding Q6, known vs. unknown actors. Younger participants show
more security/privacy concern regarding unauthorized access from both known and
unknown threats, and are particularly sensitive to known actors as compared to
other groups. 

The second factor solution (F2) also exposes an overall difference in
the security/privacy sensitivity for age. Recall that more negative responses
correlated with more security/privacy sensitivity, and we see an increase in
that sensitivity (more negative) for increased age. Looking at the
weightings to questions within F2, the response to Q1 (concern about
security/privacy), Q4 (identify threat), and Q5 (online privacy) are the most
strongly negative within the F2 solution. The result may indicate that
these security/privacy threats resonate more strongly with older participants
and that F2 is capturing that.

As a limitation to these results from age and location, there is overlap between
individual groupings in these statistical tests, e.g., more young people live in
urban settings, so results may be oversampling a sub-group of the
population. Even correcting for that, across the hypothesis (using smallest
p-values), it would suggest that Q2, Q3, and Q6 play the most significant
individual roles; this is further supported in the PCA weightings for both F1
and F2 solutions.

% \begin{figure}[t]
%   \centering
%   \includegraphics[width=1.0\linewidth]{figures/guess.pdf}
%   \caption{Scatter plot of mean privacy response to guessing number of
%     reported patterns. There is a slight positive correlation of
%     $\rho=0.08$ with $p<0.1$.}
%   \label{fig:scatter}
% \end{figure}

\subsection{Guessability Strength}

\begin{figure}[t]
  \centering
  \includegraphics[width=0.75\linewidth]{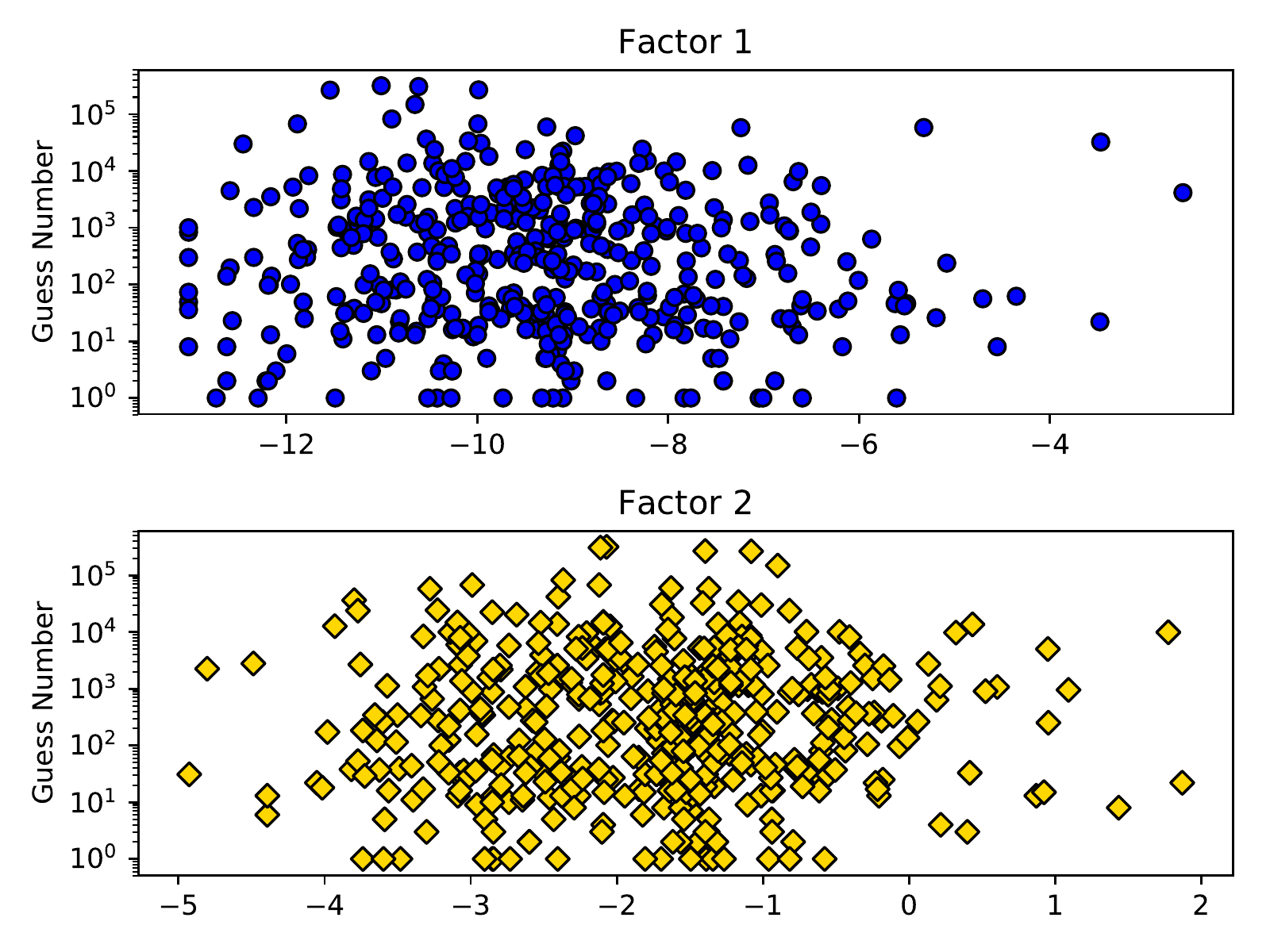}
  \caption{Scatter plot of factor solutions to the guessing number of reported
    patterns. There is a negative correlation of $r=-0.08$ using Pearson rank
    for Factor 1 with small a effect with a small, an inconclusive effect size
    $p<0.1$, leading to a potential negative result. No potentially significant
    correlation for Factor 2 was found.}
  \label{fig:scatter}
\end{figure}

In Figure~\ref{fig:scatter}, a scatter plot is presented which depicts the PCA
two factor solutions for each participant who self-reported a pattern compared
to the guessing number (log scale). Using Pearson Correlation (on the non-log
guess values), we found that there exists a small correlation between the
average security/privacy response and the guessing number ($r=-0.08, p<0.1$)
for F1. This suggests that as individuals express stronger security/privacy
preferences (more negative results in F1 indicates more agreement with the
security/privacy prompts), there is a small increase in self-reported passcode
strength, as represented in the guessing number. {\em We acknowledge that this
  is a limited and disheartening small effect and possibly inconclusive or a
  negative result, but this and other results do suggest that there likely
  exists some correlation between security/privacy attitudes and security of
  authentication choice under the right measurement circumstances.} Our
intuition for there likely being a connection is due to the responses to
individual questions with regard to the security of authentication choices.

\begin{figure}[t]
  \centering
  \includegraphics[width=0.75\linewidth]{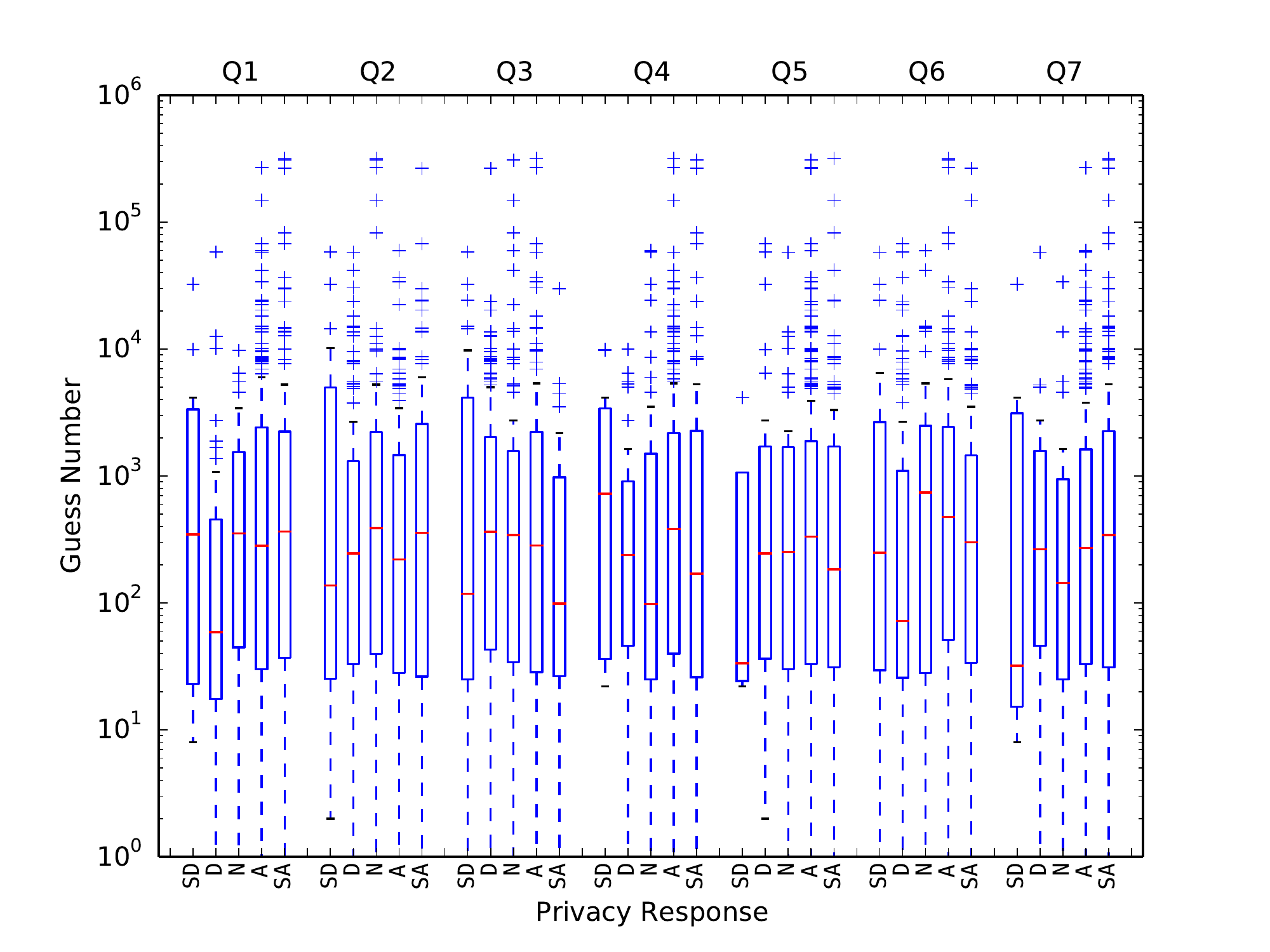}
  \caption{Box-and-Whisker plot for guessing number strength (log-scale) compared
    to security/privacy response where: SD is ``Strongly Disagree,'' D is
    ``Disagree,'' N is ``Neither Disagree nor Agree,'' A is ``Agree,''
    and SA is ``Strongly Agree.''}
  \label{fig:guess_per}
\end{figure}

Breaking down the results further per question to analyze the attitudes towards
specific security/privacy prompts that have the largest
impact. Figure~\ref{fig:guess_per} displays a box-and-whisker plot of the
guessing numbers for self-reported patterns broken down by response within each
question prompt. To measure significance, we performed a Mann-Whitney U-test
(guessing numbers are not necessarily normally distributed) within group
analysis comparing the guessing number of participants who expressed agreement
(``Agree'' or ``Strongly Agree'') with those that express disagreement
(``Disagree'' or ``Strongly Disagree''). We do not include the neutral
(``Neither Agree nor Disagree'') group in this analysis. In total, we make 7
hypothesis tests over the same overlapping data (namely guessing numbers) that
examine whether agreement/disagreement of the prompt impacts guessing number,
and as such, we consider significance with a Bonferroni
correction. Table~\ref{tab:guess_stats} displays the primary results.

We find that in the case for question prompt Q1 and Q6 there is a significant
difference between individuals who agree/disagree and the security of
self-reported Android unlock patterns; however, with Bonferroni correction with
$m=7$, a weak significance of $\alpha=0.1$ occurs when $p<0.014$ so the effect
may be limited. Recall that Q1 refers to a general concern with the security/privacy of data
stored on mobile devices, and Q2 refers to concern with unauthorized access by
known actors (e.g., friends, family, coworkers). In analysis with Q1, although the
Mann-Whitney $U$-test is compatible with unequal group sizes, it is important to
note that there were only 56 participants in the Disagree/Strongly Disagree
category as compared to 351 in the Agree/Strongly Agree category, limiting the
power of the test.  However, for Q6, there is a more balanced sample (albeit
still unequal). While the power of the test for Q6 is reduced by Bonferroni
corrections, it does suggest that there are factors in the mental model of the
threat---is it someone I know?---that could impact the authentication
choices and requires further investigation. 

%% These same two question prompts showed significant differences within
%% demographic groups, so there may be some previously as well with
%% regard to participants willingness to report their Android unlock
%% pattern as oppose to reporting significance. As such, it is likely
%% that these privacy attitudes play an important role in how individuals
%% gauge the level of security they place on unlocking their device to
%% protect the information contained therein.

\begin{table*}[!t]
\small
  \centering
  \resizebox{\linewidth}{!}{%
  \begin{tabular}{ c | c c | c c}
    & Disagree/Strongly Disaggree & Agree/Strongly Agree & $U$-Test & Cohen-$d$
    \\ \hline
Q1 & $\mu=2568.30,\sigma=8935.48,N=56$ & $\mu=7111.18,\sigma=32861.65,N=351$ & *$p=0.011$ & d=0.148\\
Q2 & $\mu=3293.11,\sigma=8799.23,N=150$ & $\mu=4609.32,\sigma=20723.04,N=201$ & $p=0.392$ & d=0.079\\
Q3 & $\mu=4218.82,\sigma=20383.17,N=184$ & $\mu=6801.41,\sigma=33295.58,N=166$ & $p=0.188$ & d=0.094\\
Q4 & $\mu=1496.86,\sigma=2688.57,N=51$ & $\mu=7028.29,\sigma=33289.51,N=339$ & $p=0.238$ & d=0.178\\
Q5 & $\mu=5974.97,\sigma=15860.97,N=32$ & $\mu=6285.81,\sigma=31447.69,N=378$ & $p=0.418$ & d=0.010\\
Q6 & $\mu=3450.97,\sigma=10158.65,N=146$ & $\mu=8277.26,\sigma=39266.21,N=235$ & *$p=0.010$ & d=0.153\\
Q7 & $\mu=3499.11,\sigma=10800.16,N=35$ & $\mu=6629.40,\sigma=31722.95,N=377$ & $p=0.296$ & d=0.102\\

    % Q1 & $\mu=2568.30,\sigma=8935.48$ & $\mu=7111.18,\sigma=32861.65$ & **$p=0.011$ & $d=0.148$\\
    % Q2 & $\mu=3293.11,\sigma=8799.23$ & $\mu=4609.32,\sigma=20723.04$ & $p=0.392$ & $d=0.079$\\
    % Q3 & $\mu=4218.82,\sigma=20383.17$ & $\mu=6801.41,\sigma=33295.58$ & $p=0.188$ & $d=0.094$\\
    % Q4 & $\mu=1496.86,\sigma=2688.57$ & $\mu=7028.29,\sigma=33289.51$ & $p=0.238$ & $d=0.178$\\
    % Q5 & $\mu=5974.97,\sigma=15860.97$ & $\mu=6285.81,\sigma=31447.69$ & $p=0.418$ & $d=0.010$\\
    % Q6 & $\mu=3450.97,\sigma=10158.65$ & $\mu=8277.26,\sigma=39266.21$ & **$p=0.010$ & $d=0.153$\\
    % Q7 & $\mu=3499.11,\sigma=10800.16$ & $\mu=6629.40,\sigma=31722.95$ & $p=0.296$ & $d=0.102$\\
  \end{tabular} }
  \caption{Significance testing for self-reported pattern strength
    based on agreement/disagreement with question prompt. Significance testing performed using Mann-Whitney $U$-test as guessing numbers are not normally distributed. The effect size was calculated considering groups of unequal size. }
  \label{tab:guess_stats}
\end{table*}

%%% Local Variables:
%%% TeX-master: "main"
%%% End:

\section{Limitations}

There are a number of limitations in this study that are worth
noting. Foremost is that the data relies on both self-reported
attitudes towards privacy prompts as well as self-reported
authentication choices. While it could be the case that individuals
are not forthcoming, the reported authentication choices are in line
with other published results~\cite{aviv2015bigger, uellenbeck,
  song2015meteres, andriotis2013pilot}.

Additionally there may be selection bias in that participants that are
more willing to participate in the survey may already be less privacy
conscious than those who do not. We attempted to control for this
effect by allowing participants to opt-out of reporting their pattern
directly, and in comparing those two groups (reported patterns
vs. stats) we did not see any differences in the responses to the
privacy prompts, suggesting that within the groups there is a
diversity of privacy attitudes. This control does not account for
measurements of individuals who chose not to participate at all and
avoided the survey due to privacy concerns.

It is acknowledged that this study focuses solely on graphical
password choices, rather than the range of unlock authentication mechanisms used on mobile devices (i.e.,
text-based password and PIN/passcodes). The focus on graphical passwords was thought to be
a reasonable choice given prior-published guessing/security metrics
exist for Android's unlock pattern that would allow for ready-made
comparisons. We do expect that similar results would likely hold
for other unlock authentication options. 

For purposes of analysis, we examined participant agreement/disagreement with statements regarding the information privacy of mobile phones, similar to that of a Westin Scale. The Westin Scale has been examined extensively by researchers in order to determine the merits of comparing self-report vs observed behavior data regarding privacy practices ~\cite{jensen2005privacy}, enabling researchers to categorize individuals into privacy groups~\cite{woodruff2014would}. While there have been some challenges identified using such scales (e.g., between the correlation between categories and behavioral intent, and between the Westin categories and consequences), the same researchers have found that segmentations derived from the results may help to reveal deeper insights into views about privacy~\cite{woodruff2014would}.  Other researchers have questioned the applicability of the scale (e.g., \cite{king2014taken}), and the limitations imposed by its unidimensional nature compared with other scales  \cite{egelman2015scaling}. However, according to \cite{kumaraguru2015privacy}, described by \cite{egelman2015scaling}, it is still a useful tool that researchers use to examine how privacy attitudes evolve over time. Further work would be undertaken to better understand the limitations of the scale.

We also acknowledge throughout that the effect size and power of many of the
statistical tests are limited. We believe that this work shows that there is a
potential connection between privacy attitude and unlock authentication
strength. There are many factors, both in the choice of the privacy scale and
the choice (and construction/training) of the guessability metric that limit the
scope of the results. Further analysis of this connection would need to be
performed to make stronger conclusions. 

Finally, there is a limitation in that this data was collected in the Fall of
2015, before the FBI and Apple incident involving the unlocking of the San
Bernardino terrorist's iPhone~\cite{iphoneNYT}. It may be the case that
individuals are more privacy sensitive if the study were re-run today given the
publicity of government surveillance, particularly to Q7. It would be
interesting future work to compare these results to a similar study conducted in
the current climate regarding the privacy of mobile devices.

%%% Local Variables:
%%% mode: latex
%%% TeX-master: "main"
%%% End:

%\vspace{-.1in}
\section{Conclusion}

%% ------Add to Discussion section
%% -People may not be forthcoming to supply patterns

%% -Graphical pattern focus

%% -iPhone scandal – see how results vary

To conclude, we analyzed a dataset collected via an online survey on
Amazon Mechanical Turk of participants agreement/disagreement with a
set of seven security/privacy prompts related to information security on mobile
devices. Analyzing the responses using principal component analysis (PCA)
revealed two factor-solutions. We further analyzed the solutions'
weighting and responses to individual questions based on demographic
breakdowns, willingness to reveal or not reveal an unlock pattern,
and the strength of the unlock pattern if revealed. 

Originally, we hypothesized that (H1) there is a relationship between
the responses to the security/privacy prompts and the strength of the reported
authentication stimuli, and (H2) that there is a relationship between
the participants' demographic information (e.g., age, location of
residence) and the responses to the security/privacy prompts. 

In regards to H1, we were unable to find sufficient evidence to accept or reject
H1 outright. We found weak evidence, potentially inconclusive evidence with
regard to one of our factor solutions (F1), in support of H1, but the effect
size and the correlation results are not significant.  Our second factor
solution (F2) showed no significant differences between those that revealed and
those that did not reveal their unlock pattern, but again, this does not allow
us to draw conclusions regarding the relationship between pattern strength and
security/privacy attitudes. Further research would be needed and perhaps a modification
to either the security/privacy prompts or the security metric, or both.

However, while we were unable to identify larger trends that include
the entire set of security/privacy prompts, we did find that individual
questions provided significant differences in the security of reported
unlock patterns. Both Q1 (concern for privacy of data on mobile
devices) and Q6 (concern about unauthorized access by known actors)
proved to divide the groups well where those who agreed more strongly
selected stronger patterns. This suggests that modifications to the
mobile device specific security/privacy prompts should include questions that
explore this connection further.

While results for H1 are inconclusive, we are able to accept H2. There are demographic differences in the responses to the security/privacy prompts,
particularly as it relates to both of our factor solutions (F1 and
F2). With respect to place of residence, when applying F1, we find
that both urban and rural participants have similar privacy responses
(more privacy conscious) that differ from sub-urban participants (less
privacy conscious). When considering age, F2 showed very strong
differences in the responses as participants got older, and based on
the weighting, we believe this relates to the responses to Q1 (concern
about privacy), Q4 (identify threat), and Q5 (online privacy) which
older participants had more agreement with.

Based on these findings, we believe that incorporating some or all of the
security/privacy prompts as a short survey within the password creation process
could help inform a system or policy makers on how to better personalize the
selection process. For example, if Android used a similar prompt during
initialization, the information from that response can nudge users who are more
casual about information security and privacy towards stronger authenticator.

%%% Local Variables:
%%% TeX-master: "main"
%%% End:

\section*{Acknowledgments}
This work was supported by the Office of Naval Research and the National
Security Agency. We would like to thank Stephen Chan, Devon Budzitowski, Jeanne
Lunning-Prak, and Justin Maguire for help supporting this research.

\bibliographystyle{plain}

\bibliography{ref}
\balance
\end{document}